\documentclass[11pt]{article}
\RequirePackage[OT1]{fontenc}
\usepackage{amsthm,amsmath,amssymb,amsfonts,bm,enumerate,xcolor,graphicx,natbib,color,url}
\RequirePackage[colorlinks,citecolor=blue,urlcolor=blue]{hyperref}
\usepackage{hyperref}
\usepackage{subcaption} 
\usepackage{ulem}
\usepackage{pifont}
\usepackage{textcmds}
\usepackage{booktabs}
\usepackage{multirow}
\usepackage{booktabs}
\usepackage{adjustbox}
\usepackage[usestackEOL]{stackengine}
\usepackage[toc,page]{appendix} 
\usepackage{algorithm,algorithmic}
\usepackage{lipsum}
\usepackage{setspace}

\usepackage{setspace}
\allowdisplaybreaks
\def\frac#1#2{{\textstyle{#1\over#2}}}

\DeclareSymbolFont{AMSb}{U}{msb}{m}{n}
\DeclareMathSymbol{\Natural}{\mathbin}{AMSb}{"4E}
\DeclareMathSymbol{\Integer}{\mathbin}{AMSb}{"5A}
\DeclareMathSymbol{\Real}{\mathbin}{AMSb}{"52}
\DeclareMathSymbol{\Rational}{\mathbin}{AMSb}{"51}
\DeclareMathSymbol{\Imaginary}{\mathbin}{AMSb}{"49}
\DeclareMathSymbol{\Complex}{\mathbin}{AMSb}{"43} 
\DeclareMathSymbol{\Disk}{\mathbin}{AMSb}{"44} 
\def\bi{\begin{itemize}}
\def\ei{\end{itemize}}
\def\bd{\begin{description}}
\def\ed{\end{description}}
\def\ben{\begin{enumerate}}
\def\een{\end{enumerate}}
\def\bv{\begin{verbatim}}
\def\ev{\end{verbatim}}

\def\pr{{\rm Pr}}
\def\Pr{\pr}

\def\2to{{\ {\buildrel 2\over \longrightarrow}\ }}

\def\I1ton{{$I_1,\ldots,I_n$}}
\def\X1ton{{$X_1,\ldots,X_n$}}
\def\Y1ton{{$Y_1,\ldots,Y_n$}}
\def\Z1ton{{$Z_1,\ldots,Z_n$}}
\def\R1ton{{$R_1,\ldots,R_n$}}
\def\e1ton{{$e_1,\ldots,e_n$}}
\def\t1ton{{$t_1,\ldots,t_n$}}
\def\x1ton{{$x_1,\ldots,x_n$}}
\def\y1ton{{$y_1,\ldots,y_n$}}
\def\z1ton{{$z_1,\ldots,z_n$}}

\newcommand{\blind}{1}

\begin{document}
\thispagestyle{empty}
\baselineskip=28pt
\vskip 5mm

\begin{center} 
{\Large{\bf Modeling of spatial extremes in environmental data science: \\ Time to move away from max-stable processes}}
\end{center}

\baselineskip=12pt

\vskip 5mm

\if1\blind
{
\begin{center}
\large
Rapha\"el Huser$^1$, Thomas Opitz$^2$, and  Jennifer Wadsworth$^3$
\end{center}
\footnotetext[1]{
\baselineskip=10pt Statistics Program, Computer, Electrical and Mathematical Sciences and Engineering (CEMSE) Division, King Abdullah University of Science and Technology (KAUST), Thuwal 23955-6900, Saudi Arabia. E-mail: raphael.huser@kaust.edu.sa}
\footnotetext[2]{
\baselineskip=10pt  INRAE, Biostatistics and Spatial Processes (BioSP, UR546), 228 Route de l'A\'erodrome, 84914 Avignon, France. E-mail: thomas.opitz@inrae.fr}
\footnotetext[3]{
\baselineskip=10pt  Department of Mathematics and Statistics, Fylde College, Lancaster University, Lancaster, United Kingdom. E-mail: j.wadsworth@lancaster.ac.uk}
}\fi

\baselineskip=26pt
\vskip 2mm
\centerline{\today}
\vskip 4mm


\baselineskip=14pt

{\large{\bf Abstract.}} Environmental data science for spatial extremes has traditionally relied heavily on max-stable processes. Even though the  popularity of these models has perhaps peaked with statisticians, they are still perceived and considered  as the `state-of-the-art' in many applied fields. However, while the asymptotic theory supporting the use of max-stable processes is mathematically rigorous and comprehensive, we think that it has also been overused, if not misused, in environmental applications, to the detriment of more purposeful and meticulously validated models. In this paper, we review the main limitations of max-stable process models, and strongly argue against their systematic use in environmental studies. Alternative solutions based on more flexible frameworks using the exceedances of variables above appropriately chosen high thresholds are discussed, and an outlook on future research is given, highlighting recommendations moving forward and the opportunities offered by hybridizing machine learning with extreme-value statistics.

\vspace{10pt}

{\large{\bf Impact statement.}} This position paper reviews the severe limitations of max-stable processes for environmental extreme data science, and discusses more appropriate alternative statistical frameworks for the modeling of spatial extremes that have emerged recently. Use of machine learning and AI methods in spatial extreme-value modeling and inference is also discussed, and seven key recommendations to push the field forward are given.

\vfill


\baselineskip=20pt

\newpage

\section{Introduction} 
\label{sec:Introduction}

With the rise of statistical machine learning that marks the `data science revolution' \citep{donoho2017}, and the increasing availability of massive high-quality environmental data products based on observation and simulation (e.g., large climate model ensembles and  reanalysis data, remote sensing, wide in-situ observation networks, mobile sensors or citizen-science data), the relevance of traditional statistical models 
is at stake more than ever. This is especially true with the modeling and prediction of 
environmental extreme events, 
where assumptions are crucial for accurate risk assessment and mitigation, and where applied findings are of key societal importance on a global scale \citep{ipcc2023}. On the one hand, sophisticated models that are well supported by probability theory are desired, in order to provide sound extrapolation into the tail of the distribution. On the other hand, practical considerations should guide the model construction to ensure that it can efficiently utilize available data and provide the answers we need to appropriately address the specific scientific problem at hand. In particular, the variability of processes along the space and time dimensions usually plays a key role in environmental science, and  a given statistical model should capture the most important 
marginal and dependence features of the data, such as spatio-temporal trends and non-stationarity, non-Gaussianity, and subasymptotic forms of tail dependence. Such models should also enable fast-enough inference, which includes model fitting, validation, simulation, and prediction. The speed at which a statistical extreme-value analysis must be performed and the amount of human and material resources required to achieve it depend strongly on the context; while spending months or years could be 
acceptable for academic purposes or for retrospective studies, it is crucial in some cases to do it within just a few days or weeks (as, e.g., with rapid extreme-event attribution studies to respond to the media about the role of anthropogenic forcings in the occurrence of a recent catastrophic event; see \citealp{Stott.etal:2004,Risser.Wehner:2017}) or even `online' (as, e.g., with operational early-warning systems predicting natural hazards in real time, where the safety of people or infrastructure is at risk; see \citealp{Nguyen.etal:2023}). Oftentimes, however, these different requirements 
are at odds with each other: popular spatial model classes arising from asymptotic extreme-value theory are often computationally prohibitive due to their intricate probabilistic structure, or subject to important practical restrictions that hamper their widespread application in real operational settings, where data are often big and complex. In this paper, we recall that an asymptotic motivation should never supersede applied scientific considerations and proper model checking. Importantly, we argue that the class of \emph{max-stable processes}, characterized by extreme-value copulas, whose practical usage by statisticians and climate scientists has been multiplied since the article of \citet{padoan2010likelihood}, not only has many severe built-in limitations, but also fails to address the basic purpose it was made for: namely, to provide a suitable statistical framework for modeling spatial extremes and estimating small joint tail probabilities (or, similarly, high return levels of spatial aggregates) far beyond observed levels. The enthusiasm about, and adoption of, max-stable process models in environmental studies---for which we are partly responsible---is due to their solid theoretical foundation, which makes them appear as `natural' models to use, the fact that the extreme-value community has traditionally been more theory-oriented while being somewhat `detached' from concrete issues arising in real applications, and the availability of convenient user-friendly software such as the \texttt{R} package \texttt{SpatialExtremes} \citep{ribatet2022SpatialExtremes} to fit and simulate these models. While (part of) the statistics of extremes community has already realized some of their limitations \citep{davison2019spatial,huser2022advances} and started to develop alternative modeling strategies that transcend the classical framework \citep{wadsworth2012dependence,opitz2016modeling,deFondevilleDavison18,HuserWadsworth19,engelke2020graphical,huser2021maxid,wadsworth2022higher,castrocamilo2022practical}, max-stable processes and extreme-value copulas still continue to be used in many spatial data applications and considered in simulation studies as the `default' option. 

This paper aims to openly discuss the known deficiencies of max-stable processes, and strongly encourage statisticians and climate scientists to move away from them in real applications unless better alternative solutions are not available. We argue that, as a community, it is now time to reflect and act upon the lessons learned over the past two decades, and move on with the broader adoption of more recent, flexible, efficient, and pragmatic solutions for the modeling of extremes in operational risk assessment studies and, more generally, environmental data science.

The rest of the paper is organized as follows. In Section~\ref{sec:MSPs}, we review max-stable processes and their main limitations. In Section~\ref{sec:Solutions}, we discuss alternative modeling strategies. In Section~\ref{sec:Conclusion}, we conclude with some final remarks, an outlook on the future of environmental extreme data science, with a particular view on advances at the interface between statistics of extremes and modern machine learning, and we also list several recommendations moving forward.


\section{Max-stable processes: a restrictive tool for the wrong problem?} 
\label{sec:MSPs}

The theoretical foundations underpinning max-stable processes start with the wish to generalize univariate extreme-value theory \citep{davison2015statistics} to the spatial context. Consider a sequence of independent and identically distributed random processes, $Y_1(\bm s),Y_2(\bm s),\ldots$, defined over a spatial domain $\mathcal{S}$. Extreme-value theory states that under broad conditions, the only possible limits, $Z(\bm s)$, for the process of pointwise maxima, $Z_n(\bm s)=\max\{Y_1(\bm s),\ldots,Y_n(\bm s)\}$ as $n\to\infty$, when appropriately affinely renormalized, are max-stable processes. This result implies that all univariate margins of $Z(\bm s)$ follow the generalized extreme-value (GEV) distribution, while all finite-dimensional margins are characterized by an extreme-value copula \citep{davison2012statistical,Segers:2012}. This asymptotic characterization (as the block size $n$ tends to infinity) has been the principal argument for fitting max-stable models and extreme-value copulas in practice (with fixed and finite $n$). While the max-stable process theory was established in the 1980s, their popularity started to grow with \citet{schlather2002models} who showed how to construct max-stable models with realistic-looking realizations  based on \citet{de1984spectral}'s spectral representation, and their use in environmental applications was later boosted by \citet{padoan2010likelihood} who proposed a method of inference for spatial max-stable models based on pairwise likelihoods. \citet{davison2012statistical} further advocated their use against other natural alternatives available at the time. The spectral representation of max-stable processes \citep{de1984spectral, schlather2002models} essentially states that, on the unit Fr\'echet scale, they can be constructed as
\begin{equation}
\label{eq:MSP}
    Z(\bm s)=\sup_{i\geq 1}\xi_i W_i(\bm s),
\end{equation}
for a Poisson point process $\{\xi_i\}_{i\geq1}$ on $(0,\infty)$ with intensity $\xi^{-2}\mbox{d}\xi$, and independent copies $W_i(\bm s)$ of a spatial process $W(\bm s)$ satisfying $\mbox{E}[\max\{W(\bm{s}), 0\}] = 1$. The unit Fr\'echet scale is a common standard since it permits simple expression of the finite-dimensional distribution functions as $G(\bm z)=\exp\{-V(\bm z)\}$, with $V$ the so-called exponent function, which is homogeneous of order $-1$, meaning that $t\times V(t\bm z) = V(\bm z)$ for any positive $t$ and $\bm z$. This construction principle has led to various max-stable process models, the most popular ones being the \citet{smith1990max} model,  the \citet{schlather2002models} model, the extremal-$t$ model \citep{opitz2013extremal}, the \citet{brown1977extreme} model \citep{kabluchko2009}, and the \citet{reich2012hierarchical} model, which have been widely used in numerous applications. However, max-stable processes have many intrinsic limitations and practical restrictions, and we now summarize
 the most critical ones. These restrictions are related to the actual definition of max-stable processes, the form of their dependence structure, and the cost of likelihood inference and simulation algorithms.

\begin{figure}
    \centering
    \includegraphics[width=0.48\linewidth]{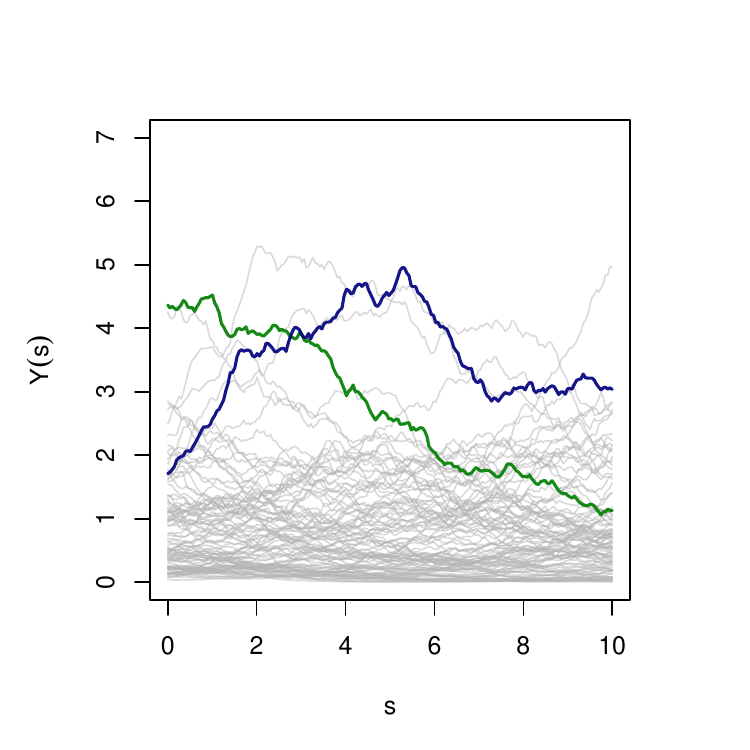}
    \includegraphics[width=0.48\linewidth]{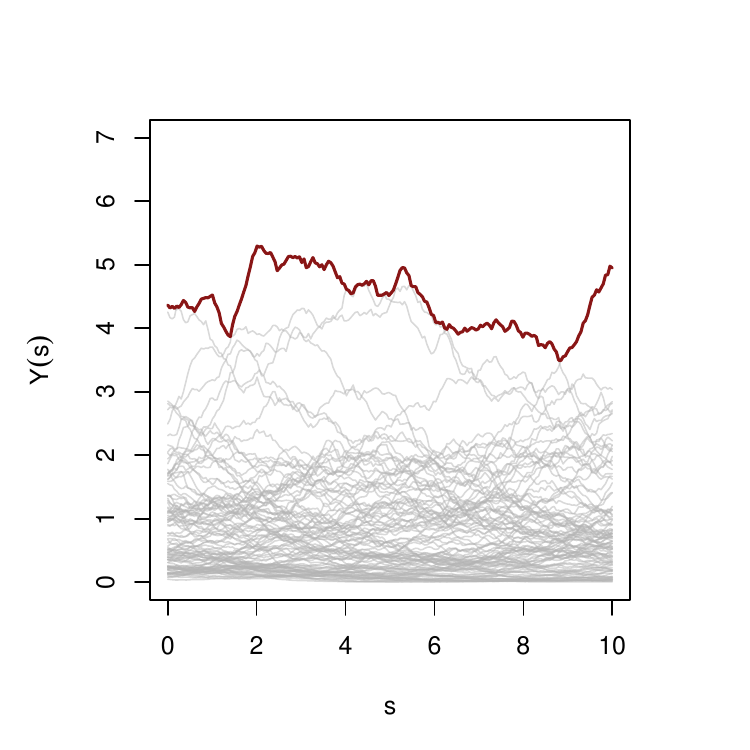}
    \caption{\emph{Left:} Illustration of underlying random fields $Y_i(\bm s)$ in gray, with two of the fields that contribute to the pointwise maximum highlighted in green and blue. \emph{Right:} Illustration of underlying random fields $Y_i(\bm s)$ in gray, with the pointwise maximum $Z_n(\bm s)$ highlighted in red. In this case, there is no $j=1,\ldots,n$ such that $Z_n(\bm s)=Y_j(\bm s)$ for all $\bm s$.}
    \label{fig:MSP}
\end{figure}

\paragraph{Definition of max-stable processes} A first drawback of max-stable processes goes back to their basic theoretical motivation, and the implicit definition of a spatial extreme event in this framework. By construction, max-stable processes $Z(\bm s)$ approximate the distribution of pointwise maxima $Z_n(\bm s)$ for large $n$. However, while the $Y_i(\bm s)$'s represent the original individual (e.g., daily) spatial events, which are directly observable and may or may not be extreme, the block (e.g., yearly) maximum process $Z_n(\bm s)$ does not correspond to a real event, unless a single individual process, say $Y_j(\bm s)$, is more extreme than \emph{all} other processes $Y_i(\bm s)$, $i\in\{1,\ldots,n\}\setminus\{j\}$, simultaneously at \emph{all} sites $\bm s$ within the domain $\mathcal{S}$, such that $Z_n(\bm s)=Y_j(\bm s)$ for all $\bm s$. This situation rarely occurs in practice, especially for large domains. Therefore, in fitting a max-stable process to realizations of $Z_n(\bm s)$, we effectively model artificially-created spatial extreme data that may have little to do with real observations. This is illustrated in Figure~\ref{fig:MSP}.
An argument used to persist and still fit max-stable processes to extremes of individual observations \citep[e.g.,][]{huser2014space} is that they provide, in some sense, an approximation to the joint tail of $Y_i$ itself. Specifically, if a vector $\bm Y\sim F$ has a joint distribution $F$ with unit Fr\'echet margins attracted to the max-stable distribution $G(\bm z)=\exp\{-V(\bm z)\}$, then we have $F(n\bm y)=[\{F(n\bm y)\}^n]^{1/n}\approx \{G(\bm y)\}^{1/n}=G(n\bm y)$, where the last equality uses the homogeneity of $V$, and the approximation is justified when $n$ is large. However, as is clear from this simple derivation, the max-stable distribution $G$ only provides an accurate approximation to $F$ when \emph{all} observations are large at the same time, and it does not provide a suitable approximation 
in the much more common situation where only a subset of variables are extreme.   
This issue can be partly dealt with by censoring small observations when performing inference; nevertheless, we argue that, by design, max-stable processes address the `wrong problem', and that the very statistical framework motivating their practical use is flawed.

\paragraph{The rigidity of  stability} Max-stable processes possess a dependence structure that is too rigid for most environmental applications. A max-stable distribution $G$ is one for which $G^t$ (with $G^t(x)$ defined as $\{G(x)\}^t$) remains a valid distribution within the same location-scale family for all $t>0$. This implies that the dependence structure of max-stable processes is invariant to the maximum operator; in other words, if a specific max-stable process is an appropriate model for annual maxima of a given variable, then the same model also provides an appropriate characterization of the dependence structure for 10-year maxima, 50-year maxima, or even 1000-year maxima. This means that, upon marginal standardization, the dependence patterns of a spatial extreme event do not 
change with the severity of the event, no matter how extreme it is. This rigidity often contradicts empirical findings, where extreme events tend to become spatially more localized as they become more extreme \citep{HuserWadsworth19,castro2020local,Zhong.etal:2021}.   Very few studies which employ max-stable processes actually scrutinize this stability property, but diagnostic tools can be found in \citet{Gabdaetal12} and \citet{huser2021maxid}, or could be adapted from available multivariate hypothesis testing tools \citep{Bucher2016}.

\paragraph{Models only for limited sub-classes of possible tail structures} The asymptotic world in which max-stable processes live is also `black and white': max-stable models are indeed always asymptotically dependent unless they are exactly independent, and they lack more nuanced representation of the important case of asymptotic independence. Asymptotic independence means that the limiting dependence structure of the normalized $Z_n(\bm s)$ corresponds to the independence copula, yet in practice there is almost always residual positive dependence present in $Z_n(\bm s)$ for finite $n$, even when the limit would be independence. By fitting a max-stable process model to maxima of asymptotically independent data, one incorrectly models this residual dependence as asymptotic dependence, leading to biased extrapolation further into the tail of the distribution.
Put differently, a max-stable model fitted to block maxima stemming from data with a weakening tail dependence structure is inevitably misspecified; the effect of this misspecification 
is that max-stable processes will capture an `average' strength of dependence and, therefore, tend to slightly underestimate the occurrence probability of joint extreme events at relatively low levels but potentially grossly overestimate them at high levels situated far beyond the observed range of data. 
Another way to define asymptotic dependence of the process $Y(\bm{s})$ is to consider the coefficient $\chi_{\{i,j\}}(u)=\pr\{Y(\bm s_i)>F_i^{-1}(u),Y(\bm s_j)>F_j^{-1}(u)\}/(1-u)$, where $Y(\bm s_i)\sim F_i$ and $Y(\bm s_j)\sim F_j$. Asymptotic dependence is present if this converges to a positive limit $\chi_{\{i,j\}}>0$ as $u\to 1$. In other words, the joint tail decay rate is proportional to the marginal tail decay rate, and any dependence at moderate levels never vanishes completely but it remains in the limiting tail. While asymptotic dependence (or independence) is a property that is difficult to verify or test in practice, models allowing for asymptotic independence often have richer tail decay rates than max-stable processes, which are strongly limited due to their focus on the possible asymptotic structures rather than the subasymptotic tail behavior. 
This difference is crucial in practice, because the flexibility of a model in its joint tail dictates extrapolations to higher levels, and thus impacts risk assessment. Being always asymptotically dependent, max-stable processes will thus have a tendency to overestimate the risk of very large joint extremes. Comparative studies advising against asymptotically dependent models for environmental risk assessment include \citet{Bortot2000} for oceanographic processes and  \citet{opitz2016modeling} and \citet{Dawkins2018} for extreme wind gusts.

\paragraph{Lack of flexible and physically realistic models} Some of the most popular max-stable models are also characterized by unphysical properties. The Schlather and extremal-$t$ max-stable models, for example, are non-ergodic, which implies that they cannot approach full independence between infinitely-distanced sites. With such models, spatial extreme events have a positive probability to be `infinitely wide' in extent since the $\chi_{\{i,j\}}$ coefficient for $Z(\bm s)$ is uniformly bounded below by a positive constant for any two locations $\bm s_i$ and $\bm s_j$ whatever their distance $\|\bm s_i-\bm s_j\|$. On the other hand, the Smith model has overly smooth realizations based on analytical spatial profiles $W_i$ in \eqref{eq:MSP}, and the original Reich--Shaby model is a noisy version of the Smith model with artificial non-stationary artefacts. The Brown--Resnick model, which is constructed from intrinsically stationary log-Gaussian processes $W_i$ in \eqref{eq:MSP}, seems to be the most physically reasonable one. Nevertheless, the very broad subclass of max-stable processes possessing a positive continuous density in all of their finite-dimensional distributions (including the Brown--Resnick model itself) 
shares the common drawback that conditional independence implies full independence \citep{papastathopoulos2016conditional}. This means that most max-stable models cannot exhibit any interesting Markov structure directly in $Z(\bm s)$ but at most in latent variables used to construct the max-stable process \citep{Amendola2022,engelke2020graphical}, which is a major impediment to leveraging such Markov structures for efficient inference and validation of max-stable models.  
This also implies that physically-meaningful stochastic partial differential equation models \citep[see, e.g.,][]{lindgren2009explicit,Bolin.Wallin:2020,Zhang.etal:2023}, which commonly lead to conditional independencies and graphical structures when discretized, are directly incompatible with max-stable processes.

\paragraph{Computational complexity} Finally, max-stable processes are also notoriously computationally cumbersome to fit and simulate from. Likelihood-based inference is especially challenging, even with advanced supercomputers \citep{castruccio2016high}, because the full likelihood function contains a number of terms that grows super-exponentially fast with the number of observed locations. Therefore, alternative inference solutions are required. Proposed approaches include pairwise likelihoods \citep{padoan2010likelihood}, higher-order composite likelihoods \citep{genton2011likelihood,huser2013composite,huser2023vecchia}, M-estimators \citep{Yuen2014,einmahl2016Mestimator}, a customized stochastic expectation-maximization (EM) algorithm \citep{huser2019full}, distributed inference through a divide-and-conquer strategy \citep{Hector2023Distributed}, or more recently neural Bayes estimators \citep{lenzi2023neural,sainsbury2023fast}. However, except for the latter which 
can be trained offline, these approaches all face a delicate trade-off between computational and statistical efficiency, and they often remain quite expensive to apply in moderate-to-high dimensions in terms of number of spatial locations. Likelihood-based inference can be simplified by including event times in the dataset \citep{StephensonTawn05}, but this can lead to bias if the number of locations is large in comparison to the number of replicates over which maxima are taken \citep{Wadsworth15}. Nonetheless, the inclusion of event times starts to mimic the paradigm of peaks-over-threshold modeling, which we argue in Section~\ref{sec:Solutions} is a more sensible and natural approach.

The stochastic representation \eqref{eq:MSP} involving an infinite number of processes over which the pointwise maximum is taken also makes simulation cumbersome. While various types of approximate and exact simulation algorithms have been developed for certain max-stable families \citep{schlather2002models,oesting2012simulation,Thibaud2015,dombry2016exact,oesting2022comparative}, conditional simulation remains computationally laborious \citep{dombry2013conditional,oesting2013conditional} and does not scale well with the dimension. To construct more ``generic'' stochastic generators for spatial extremes, it is also possible to directly exploit generative artificial-intelligence (AI) techniques from the machine learning literature, such as Generative Adversarial Networks (GANs; see, e.g., \citealp{Boulaguiem2022}) which bypass the need to specify a parametric dependence structure for extremes at the expense of completely abandoning any known structure, or Variational Autoencoders (VAEs; see, e.g., \citealp{lafon2023VAE,Zhang2023}), though these recent machine-learning-based approaches are sometimes difficult to train, are always data-hungry, and are often challenging to study from a theoretical perspective, e.g., in terms of the approximation quality of the trained generator, or its ability to accurately reproduce joint tail decay rates.  

\paragraph{Discussion} Overall, the difficulties with max-stable processes are `built-in': they are a direct consequence of their basic definition leading to the complex structure in \eqref{eq:MSP}, and the fact that they are not meant to describe extremes of the original individual events. The action of computing block maxima indeed masks information about the timings of events and temporal dependence, and specifically about co-occurrence of maxima at different spatial locations, which has implications for modeling, inference, and simulation. Max-stability arises as an `asymptotic artefact' resulting from taking the limit of block maxima as the block size $n$ goes to infinity; in practice, however, interest often lies in the original events themselves, rather than maxima. Moreover, even when the modeling of maxima may be desired, the effective block size is often quite moderate in most environmental data due to serial dependence and seasonality, which can create a severe mismatch between theory and practice. Using the limit model for data 
in this context severely constrains the form of dependence structures that can be obtained, in a way that is unrealistic in most environmental applications, while simultaneously complicating statistical inference significantly. 

In the next section, we summarize more recent modeling strategies that bypass several of the above roadblocks by going beyond max-stability, and that are thus better suited for the spatial modeling of extremes.

\section{Solutions beyond max-stability} 
\label{sec:Solutions}

In Section~\ref{sec:MSPs}, we argued strongly against the continued use of max-stable processes in all but exceptional circumstances. Here we outline the breadth of alternative models for spatially-indexed environmental data, 
highlighting their relative merits and potential drawbacks.

\begin{figure}
    \centering
    \includegraphics[width=\linewidth]{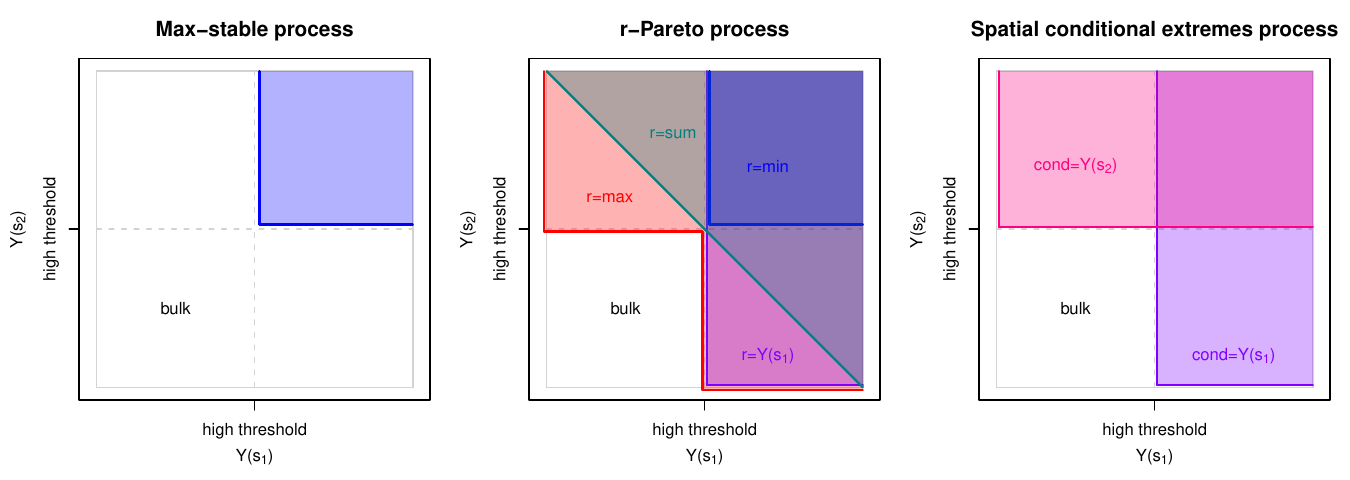}
    \caption{For a pair of variables $\{Y(\bm s_1),Y(\bm s_2)\}$ from various extreme-value models, illustration of the domains (colored areas) over which each model is meant to provide accurate tail probability approximations. \emph{Left:} max-stable process; \emph{Middle:} Pareto process for various aggregation functionals, $r$; \emph{Right:} Spatial conditional extremes process, for each conditioning variable.}
    \label{fig:maxPOT}
\end{figure}

\paragraph{Peaks-over-threshold vs maxima} Unlike approaches based on block maxima, peaks-over-threshold approaches focus on modeling extremes of the original spatial events that effectively took place. Therefore, they are not only more meaningful from a practical perspective, but they also offer ways to customize the definition of a `spatial extreme event' to the specific problem of interest. In particular, peaks-over-threshold approaches do not only provide valid probability approximations when all variables are simultaneously large (which rarely occurs in practice), but they can be adapted to events where only a subset of variables are extreme. This is illustrated in Figure~\ref{fig:maxPOT} for the two main peaks-over-threshold approaches, namely the (generalized) Pareto process \citep{FerreiradeHaan14,DombryRibatet15} and the spatial conditional extremes process \citep{wadsworth2022higher}, which are justified by different asymptotic paradigms.

Pareto processes \citep{FerreiradeHaan14,DombryRibatet15,deFondevilleDavison18,deFondevilleDavison22} are usually viewed as the peaks-over-threshold analogue of max-stable processes. Both classes of models are grounded in the theory of functional regular variation, but Pareto processes are, in principle, applicable to all data which are in some sense extreme. As a result, they possess much simpler likelihoods than max-stable models, as well as making a more efficient use of data. In Figure~\ref{fig:MSP}, Pareto processes would represent candidate models for the fields highlighted in green or blue. Slightly different formulations of Pareto processes arise depending on the so-called aggregation (or risk) functional, usually denoted by $r(\cdot)$, used to define a functional extreme event (such as the spatial maximum); see, e.g., \citet{deFondevilleDavison18,deFondevilleDavison22} for more details. 

The spatial conditional extremes model \citep{wadsworth2022higher}, on the other hand, is a different peaks-over-threshold approach that only applies to the case where a single variable (at a fixed chosen location) exceeds a high threshold; however, it offers improved tail flexibility and other benefits compared to Pareto processes, as discussed below.

\begin{figure}[t!]
    \centering
     \includegraphics[width=0.24\linewidth]{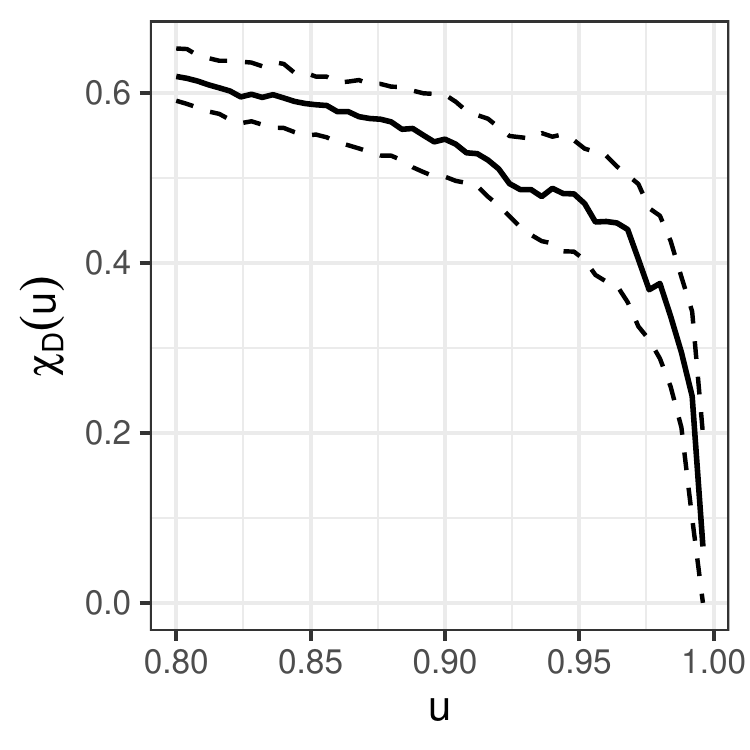}
    \includegraphics[width=0.24\linewidth]{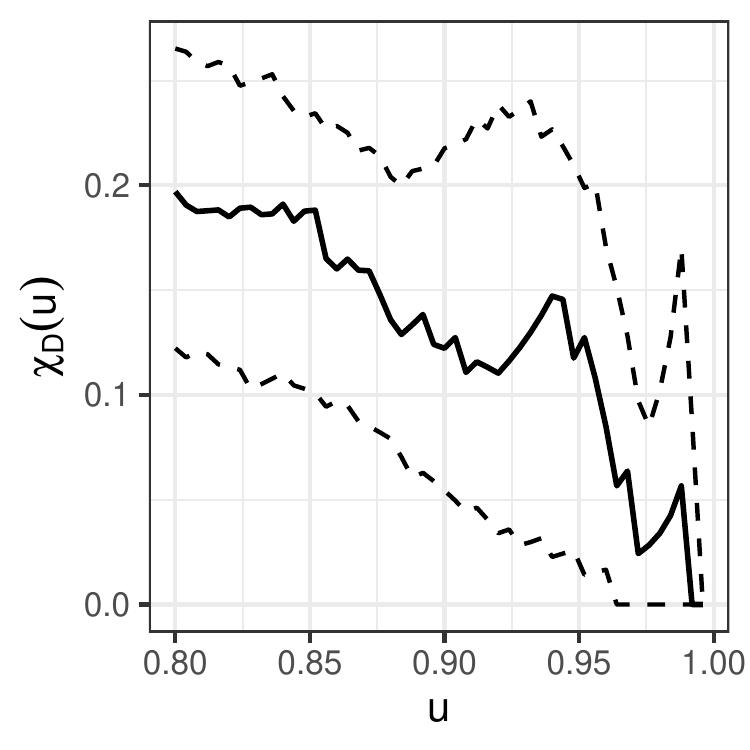}
    \includegraphics[width=0.24\linewidth]{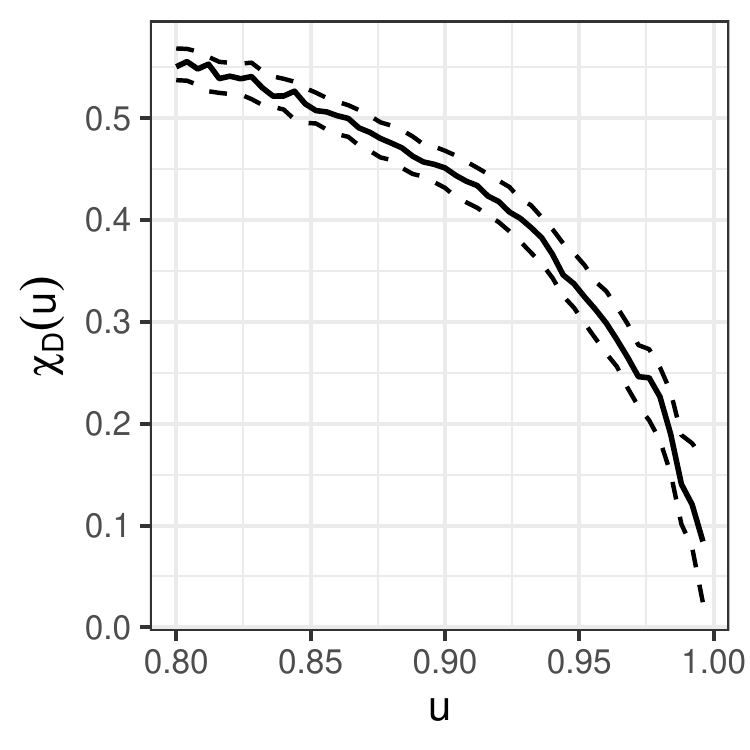}
    \includegraphics[width=0.24\linewidth]{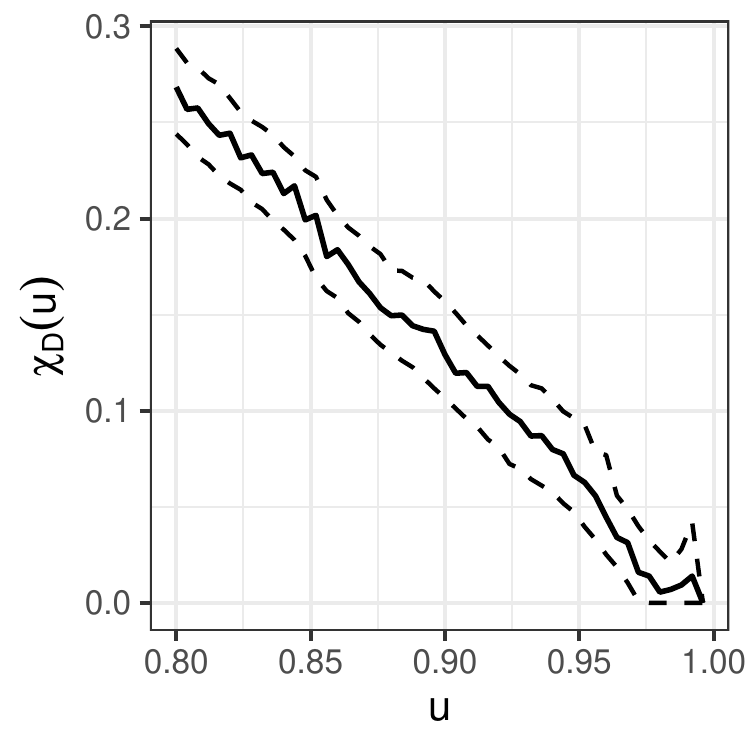}
    \caption{Examples of estimates of $\chi_D(u)$ for four environmental datasets. Solid lines represent point estimates, whereas dashed lines are approximate $95\%$ pointwise confidence intervals based on block bootstrapped estimates. From left to  right: (detrended) Red Sea surface temperature data at $d=144$ locations in the Gulf of Aqaba; gridded conditionally simulated E-OBS 
    Irish summer temperature data at $d=178$ locations; daily cumulated precipitation at $d=7$ locations in the Vaucluse `département' in France;  daily mean windspeed at the same $d=7$ locations in France.}
    \label{fig:chiu}
\end{figure}

\paragraph{Improved tail flexibility with `subasymptotic' models} While Pareto processes have simpler likelihoods and permit the use of more data than max-stable processes, they are unfortunately seldom appropriate models in practice when used for tail extrapolation and estimation of small probabilities associated with environmental extreme events that lie far in the upper joint tail. This is because their \emph{threshold-stability} property, analogous to the max-stability property of max-stable processes, is rarely satisfied by the environmental data available at observed levels of extremity. The threshold-stability property will never be satisfied by data that exhibit asymptotic independence, and represents a very strong additional assumption over the presence of asymptotic dependence. Let $Y_j = Y(\bm s_j) \sim F_j, j \in D=\{1,\ldots, d\}$, represent a spatial process observed at $d$ locations. A simple way to assess whether this property may hold is to consider the quantity $\chi_D(u)$, defined similarly to $\chi_{\{i,j\}}(u)$ in Section~\ref{sec:MSPs}, as
\begin{align}
    \chi_{D}(u) = \Pr\{Y_1>F_1^{-1}(u), \ldots, Y_d>F_d^{-1}(u)\} / (1-u), \qquad u \in (0,1).
\end{align}
If a Pareto process is applicable, then for any $d$ there always exist $d$ distinct locations such that $\chi_D(u) \equiv \chi_D >0$ for all sufficiently large $u<1$. 
Our collective experience is that, in contrast, it is almost always the case that $\chi_{D}(u)$ decreases as $u\to 1$, representing a weakening of spatial dependence at extreme levels. This is illustrated for datasets of sea surface temperature, air temperature, precipitation, and windspeed in Figure~\ref{fig:chiu}. 

To deal with this deficiency, some authors---including ourselves---have advocated the use of what are often termed \emph{subasymptotic models}. This terminology is used because such models can be seen to bridge the gap between the world of finite-level data and the `mythical land' of asymptopia, where models such as max-stable and Pareto processes should be applicable. However, it is perhaps a misnomer, since if data exhibit asymptotic independence, then Pareto processes are not well defined for domains $\mathcal{S}$ composed of an infinite number of locations, and max-stable models offer no benefits. Therefore, aside from conditional extremes models, alternatives such as these are currently the only approach to performing useful extreme value inference. 

Broadly, these subasymptotic models are designed to represent flexible forms of tail decay, permitting extrapolation from observed levels to more extreme levels. A typical consideration is that probabilities such as $\Pr\{Y_1>F_1^{-1}(u), \ldots, Y_d>F_d^{-1}(u)\}$ or $\Pr\{Y_i>F_i^{-1}(u), Y_j>F_j^{-1}(u)\}$ should have flexible forms as $u \to 1$, and these forms fit with the assumptions of regular variation and/or hidden regular variation \citep{LedfordTawn96,LedfordTawn97,Resnick02} to provide some theoretical grounding. 
Where modeling componentwise maxima remains relevant, we can relax the max-stability property to obtain more realistic dependence structures with constructions such as max-infinitely divisible processes \citep{huser2021maxid} or max-mixtures \citep{wadsworth2012dependence,Bacro2016}; however, working with such extensions of max-stable processes can further exacerbate the issues of  model interpretation and computational cost outlined before. 

To date, most subasymptotic models have a \emph{random scale} construction: extreme data are modeled using the spatial copula of the process $X(\bm s) = R W(\bm s)$, where the scalar random variable $R>0$ is independent of the spatial process $W$. The relative tail heaviness of $R$ and $W$, together with the dependence structure of $W$, provides a rich array of dependence possibilities; see \citet{Engelkeetal19} for an almost-exhaustive description. 
Examples of this class of models include \citet{opitz2016modeling}, \citet{huser2017bridging} and \citet{HuserWadsworth19}.  With appropriate specification of $R$ and $W$, Pareto processes also possess a random scale representation \citep{FerreiradeHaan14}. Limitations of simple random scale models include their inability to capture complex dependence structures observed over large domains and a gradual decay of positive dependence to independence at large distances (because of the spatially constant $R$ variable which makes them non-ergodic), as well as the incorporation of non-stationarity. These limitations also make it difficult to adapt such models to the case of spatio-temporal data.  \citet{hazra2022realistic} recently attempted to address these issues by proposing a Gaussian scale mixture model extension, which can capture short-range asymptotic dependence, mid-range asymptotic independence, and long-range exact independence, by replacing $R$ with a suitable spatial process $R(\bm s)$; see also \citet{Krupskii.Huser:2021}.

\paragraph{Conditional spatial extremes model} The recently-introduced \emph{conditional spatial extremes} model \citep{wadsworth2022higher} is another class of models with flexible tail structures, which has been adapted to the spatio-temporal case in \citet{SimpsonWadsworth21}. These models are based on the assumption of a limiting process for suitably-normalized $Y(\bm s)$, conditional upon the event $Y(\bm s_0)>t$, for some conditioning location $\bm s_0$. The formulation permits modelling of both asymptotically dependent and asymptotically independent data, while the most commonly-used version of the likelihood is relatively simple. The `price' for these two major gains is the necessity of conditioning on a particular location being large, rather than, say, \emph{any} location being large, though \citet{wadsworth2022higher} outline ways in which this can be mitigated if it is an issue.

\paragraph{Considerations for inference} As mentioned, the fitting of peaks-over-threshold models via likelihood is much simpler than max-stable models. However, to avoid bias in estimation of the tail properties, it is necessary to fit the models only to extreme data. For generalized Pareto process and subasymptotic models, this is usually done using an appropriately censored likelihood \citep[see, e.g.,][]{huser2022advances} to avoid influence of small values. Censoring can be done in different ways, either by applying a risk functional such as the maximum to the observation vector and fully censoring or discarding the vectors for which the risk is below a fixed high threshold, or using an approach focusing on marginal exceedances, without the need to define a global risk functional, where any component of the vector that falls below its marginal threshold is censored.  Especially in the latter case, this act of censoring makes likelihood evaluation significantly more computationally intensive as it can require calculating numerous potentially high-dimensional integrals of the density function. Likelihood-based inference for such models is therefore typically limited to numbers of observation locations less than about $30$, so \citet{deFondevilleDavison18} instead advocated using a weighted gradient scoring approach that mimics smooth censoring, while remaining relatively cheap computationally.

The spatial conditional extremes model can usually be fitted to data from hundreds of observation locations since its common variants take the form of a nonstationary and marginally transformed Gaussian process. The different nature of the asymptotics means that an extreme event is one that is extreme at the conditioning location $\bm s_0$, but that may be large or small elsewhere, which (often) avoids the need for censoring in the likelihood. \citet{Simpsonetal23} outline extensions to thousands of dimensions via a slight change in formulation and use of Gaussian Markov random fields; see also \citet{Vandeskog.etal:2022} for related methodology.

Techniques from the machine learning literature have recently permeated the world of spatial extreme value modeling: \citet{sainsbury2023fast}, \citet{SainsburyDaleetal23} and \citet{Walchessenetal23} describe the use of neural networks (NNs) for likelihood-free inference on the parameters of spatial models whose likelihoods are costly to evaluate because of high dimensionality. \citet{Richardsetal23} extend these ideas to incorporate censoring of small values, a key consideration for extremes. These NN-based estimators are `amortized', meaning that all computational effort is encapsulated in a training period, after which estimates can be obtained in a fraction of a second; this is in comparison to hours or sometimes days for likelihood-based estimates.



\paragraph{Discussion}  It is almost always the case that interest lies in understanding the extremes of original events, thus it makes sense to model their extremal behavior directly. This usually leads to simpler inference and more flexible classes of models. One reason sometimes cited for preferring a block maximum approach is that the resulting $Z_n(\bm s)$ should be independent in time, as the environmental events comprising the block maxima will fall in different years. Although the modeling of original events does lead to more temporal dependence in extremes, it is typically preferable to ignore this when fitting via likelihood and adjust the uncertainty of parameter estimates post-hoc \citep{FawcettWalshaw07}.

Recent developments in machine-learning-based inference begin to circumvent some computational difficulties for inference with both peaks-over-threshold models and max-stable processes alike. Nonetheless, the conceptual drawbacks of max-stable processes remain, together with potential simulation difficulties. These recent machine-learning-based developments have largely thus far focused on demonstrating how to fit existing models in previously unfeasible scenarios. An exciting possibility is their potential to facilitate generation of flexible new models with desirable tail properties: it is often easier to write down a stochastic representation for a flexible model than to derive (and evaluate) its likelihood function. For example, this could permit specification of and inference on models exhibiting asymptotic dependence at short range, with asymptotic independence and exact independence at longer range. The modeling of spatio-temporal extreme dependence, thus far tackled in relatively few cases, may also be facilitated via this route.


\section{Conclusion}
\label{sec:Conclusion}

The probability theory supporting max-stable processes is rich and should not be despised, as it has contributed significantly to extreme-value theory and led to important advances as well as a better understanding of extremes in stochastic processes; we thus do not question the rigor or historical developments of the theory itself, but rather its relevance in concrete environmental applications. 
While the systematic use of max-stable processes in environmental studies needs to be gradually phased out, we also do not categorically advocate against the use of asymptotically justified models. Asymptotic theory can indeed be very useful provided the asymptotic paradigm directly responds to a concrete need posed by the applied scientific problem at hand. In particular, peaks-over-threshold approaches should be prioritized over block maxima approaches whenever possible. Pareto processes, and the more recent spatial conditional extremes model, are two possible frameworks that stem from more helpful asymptotic regimes. However, while asymptotic guarantees are in principle desired for tail extrapolation, we also stress that they should never supersede careful model checking. Oftentimes, more pragmatic solutions (e.g., certain types of random location and scale constructions, or physics-informed models) that display improved subasymptotic tail flexibility, physically more realistic properties, or computationally more affordable inference, can be better suited to address the specific scientific problem and should thus not be disregarded. When risk assessment of future extreme events (within a reasonable time frame) is of interest, the incorporation of non-stationary climate change signals (e.g., from climate model outputs under various greenhouse gas emission scenarios) and the development of asymptotic independence models with a flexible joint tail decay rate is more important than focusing on accurately identifying the asymptotic dependence class. Furthermore, when models are intended to be used in operational settings, some accuracy must sometimes be traded for speed of inference; in this context, geostatistical models constructed from Gaussian building blocks and/or based on sparse probabilistic structures, and fast approximate inference and simulation techniques (e.g., based on deep learning), can be particularly helpful. 

An interesting direction for future research is to develop models and methods deeply rooted in extreme-value theory that harness the power and computational efficiency of advanced machine learning (e.g., deep extreme quantile regression, deep non-stationary spatial models, neural inference and generative approaches, etc.) to better address problems in environmental data science; machine learning tools can indeed overcome limitations of classical statistical tools designed for estimating few model parameters from datasets of only moderate size. \citet{Wikle.Zammit-Mangion:2023} review statistical deep learning methods in classical spatial statistics; with some suitable adjustments, these methods could potentially be adapted to the extreme-value context.  

In the future, it would also be interesting to develop unified spatial frameworks for tractably modeling the full data range in a way that offers high flexibility in the lower joint tail, bulk, and upper joint tail, in the same vein as \citet{naveau2016modeling} in the univariate context. For example, normal mean-variance mixtures allow for both asymptotic dependence and independence as well as for controlling asymmetry in lower- und upper-tail dependence \citep{Zhang2022}. Another related future line of research is to build upon recent advances in the geometric approach for multivariate extremes \citep{nolde2014,nolde2022}. The great benefit of this new asymptotic framework is that it provides a unified representation of multivariate extremes approaches \citep{wadsworth2022statistical} and a flexible strategy for the joint modeling of extremes in ``all directions''---including the lower and upper tails \citep{papastathopoulos2023statistical}; see also \citet{Murphy-Barltrop2024inference} for related methodology. Extending this approach to the spatial context and to capturing both bulk and tail behaviors simultaneously is an interesting area of investigation. 
Capturing the full range of values is particularly important for modeling compound extremes \citep{Aghakouchak2020,zscheischler2020typology} defined as the combination of conditions (over space, time or several variables) leading to extreme impacts but where  the contributing events are not necessarily extreme individually. With spatial data increasingly available on regular grids at relatively high resolution, such as climate model output and remote sensing data, the study of geometric properties of the pixellated exceedance sets at increasingly high threshold levels provides another line of research toward better understanding and modeling of the behavior of joint extremes and could offer methods that scale well to very large data volumes \citep{Cotsakis2023}. 

Although highly promising, neural likelihood-free parameter inference, mentioned in both Sections~\ref{sec:MSPs} and~\ref{sec:Solutions}, still requires several developments to replace traditional likelihood-based inference. For example, theoretical guarantees on the accuracy of neural estimators in terms of the chosen NN architecture and number of training samples remain to be established. 
Furthermore, while uncertainty quantification can be handled via the bootstrap, for example, more needs to be understood about its properties. The effects of model misspecification also need to be studied more comprehensively: with likelihood-based inference, it is known that parameter point estimates are robust to certain types of misspecification, and that adjustments can be made for properly handling uncertainty \citep{White82,ChandlerBate07}. Model selection and comparison techniques, which have often relied on log-likelihood values, also require attention. Finally, the estimation of a relatively large numbers of parameters in flexible models for spatial tail dependence remains challenging. While some of these challenges can be addressed by adopting some recent amortized fully-Bayes neural methods, such as BayesFlow \citep{radev2020bayesflow,radev2023bayesflow} or JANA \citep{radev2023jana}, their use in  extreme-value applications remains to be carefully investigated.

Our main recommendations moving forward can be summarized with the acronym MACHINE, which highlights seven `golden rules' for extreme-value analyses and the special role that machine learning is likely to play in the future of environmental extreme data science:
\begin{itemize}
\item \textbf{M}ove away from max-stable processes;
\item \textbf{A}dopt a peaks-over-threshold approach or a unified bulk-tail model whenever possible;
\item \textbf{C}apture subasymptotic behavior rather than focusing on the asymptotic structure and the dichotomy between asymptotic dependence and independence;
\item \textbf{H}arness specialized models with a sparse and numerically convenient probabilistic structure for speed and interpretation;
\item \textbf{I}ncorporate physics/climate knowledge into probability models as much as possible;
\item \textbf{N}ever-prioritize asymptotic justification over careful model checking;
\item \textbf{E}mbrace modern machine learning and AI methods to enhance the modeling and inference of extreme events in complex settings.
\end{itemize}
`Starting the MACHINE' (or keeping it on) will be key, in our opinion, to remain relevant and maximize the impact of extreme-value theory across statistics and applied environmental sciences,  especially in the face of the escalating challenges posed by today's world of extreme climate-driven events.

\section*{Acknowledgments}
We warmly thank the Environmental Data Science (EDS) editorial board and D. Nychka from the advisory board for inviting us to write this opinion piece. 

\section*{Authors contributions}
All authors contributed equally to this work. 

\section*{Competing interest statement}
The authors have no competing interests to declare.

\section*{Data availability statement}
Data to reproduce Figure~\ref{fig:chiu} can be freely obtained upon reasonable request from the authors or through the following websites: \url{https://data.marine.copernicus.eu/product/SST_GLO_SST_L4_NRT_OBSERVATIONS_010_001/description} (Red Sea surface temperature data); \url{https://www.ecad.eu/download/ensembles/download.php} (E-OBS Irish temperature data); and \url{https://meteo.data.gouv.fr/form} (French precipitation and windspeed data).

\section*{Funding statement}
This publication is based upon work supported by the King Abdullah University of Science and Technology (KAUST) Office of Sponsored Research (OSR) under Award No. OSR-CRG2020-4394.

\baselineskip 12pt
\bibliographystyle{CUP}
\bibliography{Bibliography}

\begin{thebibliography}{94}

\bibitem[AghaKouchak \emph{et~al.}(2020)AghaKouchak, Chiang, Huning, Love,
  Mallakpour, Mazdiyasni, Moftakhari, Papalexiou, Ragno and
  Sadegh]{Aghakouchak2020}
AghaKouchak, A., Chiang, F., Huning, L.~S., Love, C.~A., Mallakpour, I.,
  Mazdiyasni, O., Moftakhari, H., Papalexiou, S.~M., Ragno, E. and Sadegh, M.
  (2020) Climate extremes and compound hazards in a warming world.
\newblock \emph{Annual Review of Earth and Planetary Sciences} \textbf{48},
  519--548.

\bibitem[Am{\'e}ndola \emph{et~al.}(2022)Am{\'e}ndola, Kl{\"u}ppelberg,
  Lauritzen and Tran]{Amendola2022}
Am{\'e}ndola, C., Kl{\"u}ppelberg, C., Lauritzen, S. and Tran, N.~M. (2022)
  Conditional independence in max-linear {B}ayesian networks.
\newblock \emph{The Annals of Applied Probability} \textbf{32}, 1--45.

\bibitem[Bacro \emph{et~al.}(2016)Bacro, Gaetan and Toulemonde]{Bacro2016}
Bacro, J.-N., Gaetan, C. and Toulemonde, G. (2016) A flexible dependence model
  for spatial extremes.
\newblock \emph{Journal of Statistical Planning and Inference} \textbf{172},
  36--52.

\bibitem[Bolin and Wallin(2020)]{Bolin.Wallin:2020}
Bolin, D. and Wallin, J. (2020) Multivariate type g mat{\'e}rn stochastic
  partial differential equation random fields.
\newblock \emph{Journal of the Royal Statistical Society: Series B}
  \textbf{82}, 215--239.

\bibitem[Bortot \emph{et~al.}(2000)Bortot, Coles and Tawn]{Bortot2000}
Bortot, P., Coles, S. and Tawn, J.~A. (2000) The multivariate {G}aussian tail
  model: An application to oceanographic data.
\newblock \emph{Journal of the Royal Statistical Society: Series C (Applied
  Statistics)} \textbf{49}, 31--049.

\bibitem[Boulaguiem \emph{et~al.}(2022)Boulaguiem, Zscheischler, Vignotto,
  van~der Wiel and Engelke]{Boulaguiem2022}
Boulaguiem, Y., Zscheischler, J., Vignotto, E., van~der Wiel, K. and Engelke,
  S. (2022) Modeling and simulating spatial extremes by combining extreme value
  theory with generative adversarial networks.
\newblock \emph{Environmental Data Science} \textbf{1}, e5.

\bibitem[Brown and Resnick(1977)]{brown1977extreme}
Brown, B.~M. and Resnick, S.~I. (1977) Extreme values of independent stochastic
  processes.
\newblock \emph{Journal of Applied Probability} \textbf{14}, 732--739.

\bibitem[B{\"u}cher and Kojadinovic(2016)]{Bucher2016}
B{\"u}cher, A. and Kojadinovic, I. (2016) \emph{Extreme value modeling and risk
  analysis: Methods and applications}, chapter An overview of nonparametric
  tests of extreme-value dependence and of some related statistical procedures,
  pp. 377--398.
\newblock CRC Press.

\bibitem[Castro-Camilo and Huser(2020)]{castro2020local}
Castro-Camilo, D. and Huser, R. (2020) Local likelihood estimation of complex
  tail dependence structures, applied to us precipitation extremes.
\newblock \emph{Journal of the American Statistical Association} \textbf{115},
  1037--1054.

\bibitem[Castro-Camilo \emph{et~al.}(2022)Castro-Camilo, Huser and
  Rue]{castrocamilo2022practical}
Castro-Camilo, D., Huser, R. and Rue, H. (2022) Practical strategies for
  generalized extreme value-based regression models for extremes.
\newblock \emph{Environmetrics} \textbf{33}, e2742.

\bibitem[Castruccio \emph{et~al.}(2016)Castruccio, Huser and
  Genton]{castruccio2016high}
Castruccio, S., Huser, R. and Genton, M.~G. (2016) High-order composite
  likelihood inference for max-stable distributions and processes.
\newblock \emph{Journal of Computational and Graphical Statistics} \textbf{25},
  1212--1229.

\bibitem[Chandler and Bate(2007)]{ChandlerBate07}
Chandler, R.~E. and Bate, S. (2007) Inference for clustered data using the
  independence loglikelihood.
\newblock \emph{Biometrika} \textbf{94}, 167--183.

\bibitem[Cotsakis \emph{et~al.}(2023)Cotsakis, Di~Bernardino and
  Opitz]{Cotsakis2023}
Cotsakis, R., Di~Bernardino, E. and Opitz, T. (2023) A local statistic for the
  spatial extent of extreme threshold exceedances.
\newblock \emph{arXiv preprint arXiv:2310.09075} .

\bibitem[Davison and Huser(2015)]{davison2015statistics}
Davison, A.~C. and Huser, R. (2015) Statistics of extremes.
\newblock \emph{Annual Review of Statistics and its Application} \textbf{2},
  203--235.

\bibitem[Davison \emph{et~al.}(2019)Davison, Huser and
  Thibaud]{davison2019spatial}
Davison, A.~C., Huser, R. and Thibaud, E. (2019) Spatial extremes.
\newblock In \emph{Handbook of Environmental and Ecological Statistics}, eds
  A.~E. Gelfand, M.~Fuentes, J.~A. Hoeting and R.~L. Smith, pp. 711--744. CRC
  Press.

\bibitem[Davison \emph{et~al.}(2012)Davison, Padoan and
  Ribatet]{davison2012statistical}
Davison, A.~C., Padoan, S.~A. and Ribatet, M. (2012) Statistical modeling of
  spatial extremes.
\newblock \emph{Statistical science} \textbf{27}, 161--186.

\bibitem[Dawkins and Stephenson(2018)]{Dawkins2018}
Dawkins, L.~C. and Stephenson, D.~B. (2018) Quantification of extremal
  dependence in spatial natural hazard footprints: independence of windstorm
  gust speeds and its impact on aggregate losses.
\newblock \emph{Natural Hazards and Earth System Sciences} \textbf{18},
  2933--2949.

\bibitem[Dombry \emph{et~al.}(2016)Dombry, Engelke and
  Oesting]{dombry2016exact}
Dombry, C., Engelke, S. and Oesting, M. (2016) Exact simulation of max-stable
  processes.
\newblock \emph{Biometrika} \textbf{103}, 303--317.

\bibitem[Dombry \emph{et~al.}(2013)Dombry, \'Eyi-Minko and
  Ribatet]{dombry2013conditional}
Dombry, C., \'Eyi-Minko, F. and Ribatet, M. (2013) Conditional simulation of
  max-stable processes.
\newblock \emph{Biometrika} \textbf{100}, 111--124.

\bibitem[Dombry and Ribatet(2015)]{DombryRibatet15}
Dombry, C. and Ribatet, M. (2015) Functional regular variations, pareto
  processes and peaks over threshold.
\newblock \emph{Statistics and its Interface} \textbf{8}, 9--17.

\bibitem[Donoho(2017)]{donoho2017}
Donoho, D. (2017) 50 years of {D}ata {S}cience.
\newblock \emph{Journal of Computational and Graphical Statistics} \textbf{26},
  745--766.

\bibitem[Einmahl \emph{et~al.}(2016)Einmahl, Kiriliouk, Krajina and
  Segers]{einmahl2016Mestimator}
Einmahl, J. H.~J., Kiriliouk, A., Krajina, A. and Segers, J. (2016) An
  $m$-estimator of spatial tail dependence.
\newblock \emph{Journal of the Royal Statistical Society: Series B}
  \textbf{78}, 275--298.

\bibitem[Engelke and Hitz(2020)]{engelke2020graphical}
Engelke, S. and Hitz, A. (2020) Graphical models for extremes (with
  discussion).
\newblock \emph{Journal of the Royal Statistical Society: Series B}
  \textbf{82}, 871--932.

\bibitem[Engelke \emph{et~al.}(2019)Engelke, Opitz and
  Wadsworth]{Engelkeetal19}
Engelke, S., Opitz, T. and Wadsworth, J. (2019) Extremal dependence of random
  scale constructions.
\newblock \emph{Extremes} \textbf{22}, 623--666.

\bibitem[Fawcett and Walshaw(2007)]{FawcettWalshaw07}
Fawcett, L. and Walshaw, D. (2007) Improved estimation for temporally clustered
  extremes.
\newblock \emph{Environmetrics: The official journal of the International
  Environmetrics Society} \textbf{18}, 173--188.

\bibitem[Ferreira and {de Haan}(2014)]{FerreiradeHaan14}
Ferreira, A. and {de Haan}, L. (2014) The generalized pareto process; with a
  view towards application and simulation.
\newblock \emph{Bernoulli} \textbf{20}, 1717--1737.

\bibitem[de~Fondeville and Davison(2018)]{deFondevilleDavison18}
de~Fondeville, R. and Davison, A.~C. (2018) High-dimensional
  peaks-over-threshold inference.
\newblock \emph{Biometrika} \textbf{105}, 575--592.

\bibitem[de~Fondeville and Davison(2022)]{deFondevilleDavison22}
de~Fondeville, R. and Davison, A.~C. (2022) Functional peaks-over-threshold
  analysis.
\newblock \emph{Journal of the Royal Statistical Society Series B: Statistical
  Methodology} \textbf{84}, 1392--1422.

\bibitem[Gabda \emph{et~al.}(2012)Gabda, Towe, Wadsworth and Tawn]{Gabdaetal12}
Gabda, D., Towe, R., Wadsworth, J. and Tawn, J.~A. (2012) Discussion of
  ``statistical modeling of spatial extremes''' by a.c. davison, s.a. padoan
  and m. ribatet.
\newblock \emph{Statistical Science} \textbf{27}, 189--192.

\bibitem[Genton \emph{et~al.}(2011)Genton, Ma and Sang]{genton2011likelihood}
Genton, M.~G., Ma, Y. and Sang, H. (2011) On the likelihood function of
  gaussian max-stable processes.
\newblock \emph{Biometrika} \textbf{98}, 481--488.

\bibitem[de~Haan(1984)]{de1984spectral}
de~Haan, L. (1984) A spectral representation for max-stable processes.
\newblock \emph{The Annals of Probability} \textbf{12}, 1194--1204.

\bibitem[Hazra \emph{et~al.}(2022)Hazra, Huser and Bolin]{hazra2022realistic}
Hazra, A., Huser, R. and Bolin, D. (2022) Efficient modeling of spatial
  extremes over large geographical domains.
\newblock {arXiv preprint arXiv:2112.10248}.

\bibitem[Hector and Reich(2023)]{Hector2023Distributed}
Hector, E.~C. and Reich, B.~J. (2023) Distributed inference for spatial
  extremes modeling in high dimensions.
\newblock \emph{Journal of the American Statistical Association} To appear.

\bibitem[Huser and Davison(2013)]{huser2013composite}
Huser, R. and Davison, A.~C. (2013) Composite likelihood estimation for the
  {B}rown--{R}esnick process.
\newblock \emph{Biometrika} \textbf{100}, 511--518.

\bibitem[Huser and Davison(2014)]{huser2014space}
Huser, R. and Davison, A.~C. (2014) Space-time modelling of extreme events.
\newblock \emph{Journal of the Royal Statistical Society: Series B}
  \textbf{76}, 439--461.

\bibitem[Huser \emph{et~al.}(2019)Huser, Dombry, Ribatet and
  Genton]{huser2019full}
Huser, R., Dombry, C., Ribatet, M. and Genton, M.~G. (2019) Full likelihood
  inference for max-stable data.
\newblock \emph{Stat} \textbf{8}, e218.

\bibitem[Huser \emph{et~al.}(2017)Huser, Opitz and Thibaud]{huser2017bridging}
Huser, R., Opitz, T. and Thibaud, E. (2017) Bridging asymptotic independence
  and dependence in spatial extremes using {G}aussian scale mixtures.
\newblock \emph{Spatial Statistics} \textbf{21}, 166--186.

\bibitem[Huser \emph{et~al.}(2021)Huser, Opitz and Thibaud]{huser2021maxid}
Huser, R., Opitz, T. and Thibaud, E. (2021) Max-infinitely divisible models and
  inference for spatial extremes.
\newblock \emph{Scandinavian Journal of Statistics} \textbf{48}, 321--348.

\bibitem[Huser \emph{et~al.}(2023)Huser, Stein and Zhong]{huser2023vecchia}
Huser, R., Stein, M.~L. and Zhong, P. (2023) Vecchia likelihood approximation
  for accurate and fast inference in intractable spatial max-stable models.
\newblock \emph{Journal of Computational and Graphical Statistics} To appear.

\bibitem[Huser and Wadsworth(2019)]{HuserWadsworth19}
Huser, R. and Wadsworth, J.~L. (2019) Modeling spatial processes with unknown
  extremal dependence class.
\newblock \emph{Journal of the American statistical association} \textbf{114},
  434--444.

\bibitem[Huser and Wadsworth(2022)]{huser2022advances}
Huser, R. and Wadsworth, J.~L. (2022) Advances in statistical modeling of
  spatial extremes.
\newblock \emph{Wiley Interdisciplinary Reviews: Computational Statistics}
  \textbf{14}, e1537.

\bibitem[IPCC(2023)]{ipcc2023}
IPCC (2023) {Climate Change 2023: Synthesis Report}.
\newblock In \emph{Contribution of Working Groups I, II, and III to the Sixth
  Assessment Report of the Intergovernmental Panel on Climate Change}, eds
  H.~L. Core Writing~Team and J.~Romero, pp. 35--115. IPCC, Geneva,
  Switzerland.

\bibitem[Kabluchko \emph{et~al.}(2009)Kabluchko, Schlather and
  de~Haan]{kabluchko2009}
Kabluchko, Z., Schlather, M. and de~Haan, L. (2009) {Stationary max-stable
  fields associated to negative definite functions}.
\newblock \emph{The Annals of Probability} \textbf{37}, 2042--2065.

\bibitem[Krupskii and Huser(2022)]{Krupskii.Huser:2021}
Krupskii, P. and Huser, R. (2022) {Modeling spatial tail dependence with Cauchy
  convolution processes}.
\newblock \emph{Electronic Journal of Statistics} \textbf{16}, 6135--6174.

\bibitem[Lafon \emph{et~al.}(2023)Lafon, Naveau and Fablet]{lafon2023VAE}
Lafon, N., Naveau, P. and Fablet, R. (2023) A vae approach to sample
  multivariate extremes.
\newblock arXiv preprint arXiv:2306.10987.

\bibitem[Ledford and Tawn(1996)]{LedfordTawn96}
Ledford, A.~W. and Tawn, J.~A. (1996) Statistics for near independence in
  multivariate extreme values.
\newblock \emph{Biometrika} \textbf{83}, 169--187.

\bibitem[Ledford and Tawn(1997)]{LedfordTawn97}
Ledford, A.~W. and Tawn, J.~A. (1997) Modelling dependence within joint tail
  regions.
\newblock \emph{Journal of the Royal Statistical Society: Series B (Statistical
  Methodology)} \textbf{59}, 475--499.

\bibitem[Lenzi \emph{et~al.}(2023)Lenzi, Bessac, Rudi and
  Stein]{lenzi2023neural}
Lenzi, A., Bessac, J., Rudi, J. and Stein, M.~L. (2023) Neural networks for
  parameter estimation in intractable models.
\newblock \emph{Computational Statistics and Data Analysis} \textbf{185},
  107762.

\bibitem[Lindgren \emph{et~al.}(2009)Lindgren, Rue and
  Lindstr\"om]{lindgren2009explicit}
Lindgren, F., Rue, H. and Lindstr\"om, J. (2009) An explicit link between
  gaussian fields and gaussian markov random fields: the stochastic partial
  differential equation approach (with discussion).
\newblock \emph{Journal of the Royal Statistical Society: Series B}
  \textbf{73}, 423--498.

\bibitem[Murphy-Barltrop \emph{et~al.}(2024)Murphy-Barltrop, Mackay and
  Jonathan]{Murphy-Barltrop2024inference}
Murphy-Barltrop, C. J.~R., Mackay, E. and Jonathan, P. (2024) Inference for
  multivariate extremes via a semi-parametric angular-radial model.
\newblock arXiv preprint arXiv:2401.07259.

\bibitem[Naveau \emph{et~al.}(2016)Naveau, Huser, Ribereau and
  Hannart]{naveau2016modeling}
Naveau, P., Huser, R., Ribereau, P. and Hannart, A. (2016) Modeling jointly
  low, moderate and heavy rainfall intensities without a threshold selection.
\newblock \emph{Water Resources Research} \textbf{52}, 2753--2769.

\bibitem[Nguyen \emph{et~al.}(2023)Nguyen, Veraart, Taisne, Tan and
  Lallemant]{Nguyen.etal:2023}
Nguyen, M., Veraart, A. E.~D., Taisne, B., Tan, C.~T. and Lallemant, D. (2023)
  A dynamic extreme value model with application to volcanic eruption
  forecasting.
\newblock \emph{Mathematical Geosciences} to appear.

\bibitem[Nolde(2014)]{nolde2014}
Nolde, N. (2014) Geometric interpretation of the residual dependence
  coefficient.
\newblock \emph{Journal of Multivariate Analysis} \textbf{123}, 85--95.

\bibitem[Nolde and Wadsworth(2022)]{nolde2022}
Nolde, N. and Wadsworth, J.~L. (2022) Linking representations for multivariate
  extremes via a limit set.
\newblock \emph{Advances in Applied Probability} \textbf{54}, 688--717.

\bibitem[Oesting \emph{et~al.}(2012)Oesting, Kabluchko and
  Schlather]{oesting2012simulation}
Oesting, M., Kabluchko, Z. and Schlather, M. (2012) Simulation of
  {B}rown--{R}esnick processes.
\newblock \emph{Extremes} \textbf{15}, 89--107.

\bibitem[Oesting and Schlather(2013)]{oesting2013conditional}
Oesting, M. and Schlather, M. (2013) Conditional sampling for max-stable
  processes with a mixed moving maxima representation.
\newblock \emph{Extremes} \textbf{17}, 157--192.

\bibitem[Oesting and Strokorb(2022)]{oesting2022comparative}
Oesting, M. and Strokorb, K. (2022) A comparative tour through the simulation
  algorithms for max-stable processes.
\newblock \emph{Statistical Science} \textbf{37}, 42--63.

\bibitem[Opitz(2013)]{opitz2013extremal}
Opitz, T. (2013) Extremal t processes: Elliptical domain of attraction and a
  spectral representation.
\newblock \emph{Journal of Multivariate Analysis} \textbf{122}, 409--413.

\bibitem[Opitz(2016)]{opitz2016modeling}
Opitz, T. (2016) Modeling asymptotically independent spatial extremes based on
  {L}aplace random fields.
\newblock \emph{Spatial Statistics} \textbf{16}, 1--18.

\bibitem[Padoan \emph{et~al.}(2010)Padoan, Ribatet and
  Sisson]{padoan2010likelihood}
Padoan, S.~A., Ribatet, M. and Sisson, S.~A. (2010) Likelihood-based inference
  for max-stable processes.
\newblock \emph{Journal of the American Statistical Association} \textbf{105},
  263--277.

\bibitem[Papastathopoulos \emph{et~al.}(2023)Papastathopoulos, de~Monte,
  Campbell and Rue]{papastathopoulos2023statistical}
Papastathopoulos, I., de~Monte, L., Campbell, R. and Rue, H. (2023) Statistical
  inference for radially-stable generalized pareto distributions and return
  level-sets in geometric extremes.
\newblock arXiv preprint arXiv:2310.06130.

\bibitem[Papastathopoulos and Strokorb(2016)]{papastathopoulos2016conditional}
Papastathopoulos, I. and Strokorb, K. (2016) Conditional independence among
  max-stable laws.
\newblock \emph{Statistics and Probability Letters} \textbf{108}, 9--15.

\bibitem[Radev \emph{et~al.}(2020)Radev, Mertens, Voss, Ardizzone and
  K{\"o}the]{radev2020bayesflow}
Radev, S.~T., Mertens, U.~K., Voss, A., Ardizzone, L. and K{\"o}the, U. (2020)
  {BayesFlow}: Learning complex stochastic models with invertible neural
  networks.
\newblock \emph{IEEE transactions on neural networks and learning systems}
  \textbf{33}, 1452--1466.

\bibitem[Radev \emph{et~al.}(2023{a})Radev, Schmitt, Pratz, Picchini, Koethe
  and Buerkner]{radev2023jana}
Radev, S.~T., Schmitt, M., Pratz, V., Picchini, U., Koethe, U. and Buerkner,
  P.-C. (2023{a}) {JANA}: Jointly amortized neural approximation of complex
  {B}ayesian models.
\newblock In \emph{The 39th Conference on Uncertainty in Artificial
  Intelligence}.

\bibitem[Radev \emph{et~al.}(2023{b})Radev, Schmitt, Schumacher,
  Elsem\"{u}ller, Pratz, Sch\"{a}lte, K\"{o}the and
  B\"{u}rkner]{radev2023bayesflow}
Radev, S.~T., Schmitt, M., Schumacher, L., Elsem\"{u}ller, L., Pratz, V.,
  Sch\"{a}lte, Y., K\"{o}the, U. and B\"{u}rkner, P.-C. (2023{b}) {BayesFlow}:
  Amortized {B}ayesian workflows with neural networks.
\newblock arXiv preprint arXiv:2306.16015.

\bibitem[Reich and Shaby(2012)]{reich2012hierarchical}
Reich, B.~J. and Shaby, B.~A. (2012) A hierarchical max-stable spatial model
  for extreme precipitation.
\newblock \emph{The Annals of Applied Statistics} \textbf{6}, 1430.

\bibitem[Resnick(2002)]{Resnick02}
Resnick, S. (2002) Hidden regular variation, second order regular variation and
  asymptotic independence.
\newblock \emph{Extremes} \textbf{5}, 303--336.

\bibitem[Ribatet(2022)]{ribatet2022SpatialExtremes}
Ribatet, M. (2022) \emph{SpatialExtremes: Modelling Spatial Extremes}.
\newblock R package version 2.1-0.
  \url{https://CRAN.R-project.org/package=SpatialExtremes}.

\bibitem[Richards \emph{et~al.}(2023)Richards, Sainsbury-Dale, Zammit-Mangion
  and Huser]{Richardsetal23}
Richards, J., Sainsbury-Dale, M., Zammit-Mangion, A. and Huser, R. (2023)
  Likelihood-free neural bayes estimators for censored peaks-over-threshold
  models.
\newblock arXiv preprint arXiv:2306.15642.

\bibitem[Risser and Wehner(2017)]{Risser.Wehner:2017}
Risser, M.~D. and Wehner, M.~F. (2017) Attributable human-induced changes in
  the likelihood and magnitude of the observed extreme precipitation during
  {H}urricane {H}arvey.
\newblock \emph{Geophysical Research Letters} \textbf{44}, 12457--12464.

\bibitem[Sainsbury-Dale \emph{et~al.}(2023{a})Sainsbury-Dale, Richards,
  Zammit-Mangion and Huser]{SainsburyDaleetal23}
Sainsbury-Dale, M., Richards, J., Zammit-Mangion, A. and Huser, R. (2023{a})
  Neural {B}ayes estimators for irregular spatial data using graph neural
  networks.
\newblock arXiv preprint arXiv:2310.02600.

\bibitem[Sainsbury-Dale \emph{et~al.}(2023{b})Sainsbury-Dale, Zammit-Mangion
  and Huser]{sainsbury2023fast}
Sainsbury-Dale, M., Zammit-Mangion, A. and Huser, R. (2023{b}) Likelihood-free
  parameter estimation with neural {B}ayes estimators.
\newblock \emph{The American Statistician} To appear.

\bibitem[Schlather(2002)]{schlather2002models}
Schlather, M. (2002) Models for stationary max-stable random fields.
\newblock \emph{Extremes} \textbf{5}, 33--44.

\bibitem[Segers(2012)]{Segers:2012}
Segers, J. (2012) {Max-stable models for multivariate extremes}.
\newblock \emph{{REVSTAT}} \textbf{10}, 61--82.

\bibitem[Simpson \emph{et~al.}(2023)Simpson, Opitz and
  Wadsworth]{Simpsonetal23}
Simpson, E.~S., Opitz, T. and Wadsworth, J.~L. (2023) High-dimensional modeling
  of spatial and spatio-temporal conditional extremes using {INLA} and
  {G}aussian {M}arkov random fields.
\newblock \emph{Extremes} pp. 1--45.

\bibitem[Simpson and Wadsworth(2021)]{SimpsonWadsworth21}
Simpson, E.~S. and Wadsworth, J.~L. (2021) Conditional modelling of
  spatio-temporal extremes for {R}ed {S}ea surface temperatures.
\newblock \emph{Spatial Statistics} \textbf{41}, 100482.

\bibitem[Smith(1990)]{smith1990max}
Smith, R.~L. (1990) Max-stable processes and spatial extremes.
\newblock Unpublished manuscript.
  \url{https://www.rls.sites.oasis.unc.edu/postscript/rs/spatex.pdf}.

\bibitem[Stephenson and Tawn(2005)]{StephensonTawn05}
Stephenson, A. and Tawn, J. (2005) Exploiting occurrence times in likelihood
  inference for componentwise maxima.
\newblock \emph{Biometrika} \textbf{92}, 213--227.

\bibitem[Stott \emph{et~al.}(2004)Stott, Stone and Allen]{Stott.etal:2004}
Stott, P.~A., Stone, D.~A. and Allen, M.~R. (2004) Human contribution to the
  european heatwave of 2003.
\newblock \emph{Nature} \textbf{432}, 610--614.

\bibitem[Thibaud and Opitz(2015)]{Thibaud2015}
Thibaud, E. and Opitz, T. (2015) Efficient inference and simulation for
  elliptical {P}areto processes.
\newblock \emph{Biometrika} \textbf{102}, 855--870.

\bibitem[Vandeskog \emph{et~al.}(2022)Vandeskog, Martino and
  Huser]{Vandeskog.etal:2022}
Vandeskog, S., Martino, S. and Huser, R. (2022) {An efficient workflow for
  modelling high-dimensional spatial extremes}.
\newblock {arXiv preprint arXiv:2210.00760}.

\bibitem[Wadsworth and Campbell(2022)]{wadsworth2022statistical}
Wadsworth, J. and Campbell, R. (2022) Statistical inference for multivariate
  extremes via a geometric approach.
\newblock \emph{arXiv preprint arXiv:2208.14951} .

\bibitem[Wadsworth(2015)]{Wadsworth15}
Wadsworth, J.~L. (2015) On the occurrence times of componentwise maxima and
  bias in likelihood inference for multivariate max-stable distributions.
\newblock \emph{Biometrika} \textbf{102}, 705--711.

\bibitem[Wadsworth and Tawn(2012)]{wadsworth2012dependence}
Wadsworth, J.~L. and Tawn, J.~A. (2012) Dependence modelling for spatial
  extremes.
\newblock \emph{Biometrika} \textbf{99}, 253--272.

\bibitem[Wadsworth and Tawn(2022)]{wadsworth2022higher}
Wadsworth, J.~L. and Tawn, J.~A. (2022) Higher-dimensional spatial extremes via
  single-site conditioning.
\newblock \emph{Spatial Statistics} \textbf{51}, 100677.

\bibitem[Walchessen \emph{et~al.}(2023)Walchessen, Lenzi and
  Kuusela]{Walchessenetal23}
Walchessen, J., Lenzi, A. and Kuusela, M. (2023) Neural likelihood surfaces for
  spatial processes with computationally intensive or intractable likelihoods.
\newblock arXiv preprint arXiv:2305.04634.

\bibitem[White(1982)]{White82}
White, H. (1982) Maximum likelihood estimation of misspecified models.
\newblock \emph{Econometrica: Journal of the econometric society} pp. 1--25.

\bibitem[Wikle and Zammit-Mangion(2023)]{Wikle.Zammit-Mangion:2023}
Wikle, C. and Zammit-Mangion, A. (2023) Statistical deep learning for spatial
  and spatiotemporal data.
\newblock \emph{Annual Review of Statistics and Its Application} \textbf{10},
  247--270.

\bibitem[Yuen and Stoev(2014)]{Yuen2014}
Yuen, R. and Stoev, S. (2014) {CRPS M-estimation for max-stable models}.
\newblock \emph{Extremes} \textbf{17}, 387--410.

\bibitem[Zhang \emph{et~al.}(2023{a})Zhang, Ma, Wikle and Huser]{Zhang2023}
Zhang, L., Ma, X., Wikle, C.~K. and Huser, R. (2023{a}) Flexible and efficient
  spatial extremes emulation via variational autoencoders.
\newblock arXiv preprint arXiv:2307.08079.

\bibitem[Zhang \emph{et~al.}(2023{b})Zhang, Bolin, Engelke and
  Huser]{Zhang.etal:2023}
Zhang, Z., Bolin, D., Engelke, S. and Huser, R. (2023{b}) Extremal dependence
  of moving average processes driven by exponential-tailed {L}\'evy noise.
\newblock arXiv preprint arXiv:2307.15796.

\bibitem[Zhang \emph{et~al.}(2022)Zhang, Huser, Opitz and Wadsworth]{Zhang2022}
Zhang, Z., Huser, R., Opitz, T. and Wadsworth, J. (2022) Modeling spatial
  extremes using normal mean-variance mixtures.
\newblock \emph{Extremes} \textbf{25}, 175--197.

\bibitem[Zhong \emph{et~al.}(2022)Zhong, Huser and Opitz]{Zhong.etal:2021}
Zhong, P., Huser, R. and Opitz, T. (2022) {Modeling non-stationary temperature
  maxima based on extremal dependence changing with event magnitude}.
\newblock \emph{{Annals of Applied Statistics}} \textbf{16}, 272--299.

\bibitem[Zscheischler \emph{et~al.}(2020)Zscheischler, Martius, Westra,
  Bevacqua, Raymond, Horton \emph{et~al.}]{zscheischler2020typology}
Zscheischler, J., Martius, O., Westra, S., Bevacqua, E., Raymond, C., Horton,
  R.~M. \emph{et~al.} (2020) A typology of compound weather and climate events.
\newblock \emph{Nature Reviews Earth \& Environment} \textbf{1}, 333--347.

\end{thebibliography}

\end{document}